\documentclass[aps,prb,twocolumn,showpacs,groupedaddress,preprintnumbers]{revtex4-1}

\usepackage{graphicx}

\begin{document}

\preprint{\textit{Published version of this paper can be found at} Physica C \textbf{470}, 689 (2010).}

\title{Anisotropic magnetotransport of superconducting and normal state in an electron-doped Nd$_{1.85}$Ce$_{0.15}$CuO$_{4-\delta}$ single crystal}


%\author{}
%\thanks{}

\author{Yue Wang$^{1}$}
\email[]{yue.wang@pku.edu.cn}

\author{Hong Gao$^2$}

\affiliation{$^1$State Key Laboratory for Mesoscopic Physics and School of Physics, Peking University, Beijing 100871, People's Republic of China}

\affiliation{$^2$National Laboratory for Superconductivity, Institute of Physics and Beijing National Laboratory for Condensed Matter Physics, Chinese Academy of Sciences, Beijing 100190, People's Republic of China}


\date{\today}

\begin{abstract}
The anisotropic properties of an optimally doped Nd$_{1.85}$Ce$_{0.15}$CuO$_{4-\delta}$ single crystal have been studied both below and above the critical temperature $T_c$ via the resistivity measurement in magnetic field $H$ up to 12~T. By scaling the conductivity fluctuation around the superconducting transition, the upper critical field $H_{c2}(T)$ has been determined for field parallel to the $c$-axis or to the basal $ab$-plane. The anisotropy factor $\gamma=H_{c2}^{\parallel ab}/H_{c2}^{\parallel c}$ is estimated to be about 8. In the normal state ($50\leq T\leq180$~K), the magnetoresistance (MR) basically follows an $H^2$ dependence and for $H\parallel c$ it is almost 10 times larger than that for $H\parallel ab$. Comparing with hole-doped cuprates it suggests that the optimally doped Nd$_{1.85}$Ce$_{0.15}$CuO$_{4-\delta}$ cuprate superconductor has a moderate anisotropy.
\end{abstract}


\maketitle

%\section{}
%\subsection{}
%\subsubsection{}

% If in two-column mode, this environment will change to single-column
% format so that long equations can be displayed. Use
%\begin{widetext}
%\end{widetext}

\section{INTRODUCTION}

One of essential features of most high-$T_c$ cuprates is their quasi-two-dimensional crystal structure with CuO$_2$ layer as a key structural unit.~\cite{Orenstein00} Under this circumstance, thermodynamic and transport properties of a given cuprate superconductor usually exhibit the anisotropy when measuring along the crystallographic $c$-axis or the CuO$_2$ plane ($ab$-plane). Anisotropy factors in these fundamental physical properties are therefore crucial to know, not only for characterizing or evaluating the sample but also as important parameters to be used in theoretical models to describe high-$T_c$ superconductivity and search for its mechanism.

In many cases, however, determination of anisotropy factors is not an easy task. A well known example is to determine the anisotropy in the upper critical field $H_{c2}$, i.e., $\gamma=H_{c2}^{\parallel ab}/H_{c2}^{\parallel c}$, where $H_{c2}^{\parallel ab}$ and $H_{c2}^{\parallel c}$ are the $H_{c2}$ along $ab$-plane and $c$-axis respectively. For most high-$T_c$ cuprates the $H_{c2}$ is exceptionally large and its evaluation is limited by laboratory accessible magnetic fields $H$ and complicated by some issues such as superconducting fluctuations, especially for the less explored $H\parallel ab$-plane ($H\parallel ab$) case. Despite this fact, continuous efforts have been devoted to extracting this basic parameter, through resistive transport, magnetization, and other kinds of experiments.~\cite{Iye88,Welp89,Farrell89,Panagopoulos03,Nagasao08}

Comparing with hole-doped counterparts, anisotropic properties of electron-doped cuprates Ln$_{2-x}$Ce$_x$CuO$_{4-\delta}$ (Ln = Nd, Pr, ...) have been less studied and controversial results exist. For instance, an early work in Nd$_{2-x}$Ce$_x$CuO$_{4-\delta}$ (NCCO) reported a large $\gamma$ of 21,~\cite{Hidaka89} while subsequent magnetization measurement in aligned Sm$_{1.85}$Ce$_{0.15}$CuO$_{4-\delta}$ (SCCO) powders gave a low $\gamma$ of 3.7 and suggested that previous report might be an overestimate due to the inaccuracy in determining the $H_{c2}^{\parallel ab}$.~\cite{Almasan92} Moreover, it should be noted that in some reports a large $\gamma$ ($\sim30-200$) has been cited for describing and determining the vortex phase diagram of NCCO.~\cite{Giller97,Nugroho99} In view of these, it is desirable to redetermine the $\gamma$ in NCCO with the help of high quality crystals and proper data analysis. In this paper we report the anisotropic magnetotransport of an optimally doped NCCO single crystal. We succeeded in scaling the conductivity fluctuation near the superconducting transition in different magnetic fields for both $H\parallel ab$ and $H\parallel c$-axis ($H\parallel c$) and this enabled us to obtain $\gamma\simeq7.5$. Moreover, we also determined the anisotropy in the normal state transverse magnetoresistance (MR). Most previous MR measurements have been confined in the $H\parallel c$ configuration.~\cite{Seng95,Gollnik98} Inclusion of the data with $H\parallel ab$ helped us to confirm that the transverse MR in normal state of optimal-doped NCCO with $H\parallel c$ mainly originates from an orbital effect, namely, the bending of charge carrier trajectories due to the presence of magnetic field.

\section{EXPERIMENT}

The optimally doped Nd$_{1.85}$Ce$_{0.15}$CuO$_{4-\delta}$ single crystal was prepared by traveling solvent floating-zone technique. Resistivity measurements were carried out by the standard four-probe method with dc current supplied in the $ab$-plane. Inset of Fig. \ref{figure1} shows the temperature dependence of the resistivity, $\rho(T)$, in zero field. It is found that the crystal shows a sharp superconducting transition with an onset point at 26.2~K and a transition width around 0.7~K.

Measurements were performed in an Oxford cryogenic system (Maglab-12) with $H$ up to 12~Tesla. To determine the anisotropy, we have carefully aligned the crystal to be in $H\parallel c$ or $H\parallel ab$ configuration, by rotating the sample at fixed $H$ and $T$ in the superconducting state with an angle resolution of $1^\circ$ and tracing the peaks or minima in the angular dependence of the resistivity. The superconducting transition in magnetic fields was measured by sweeping temperature at constant $H$, while normal state MR was done by sweeping magnetic fields at fixed $T$ which was stabilized within $\sim10$~mK by a Lakeshore cernox sensor. For both field directions, the normal state MR was measured with $H$ perpendicular to the current ($H\perp I$), that is, the transverse MR was obtained.

\section{RESULTS and DISCUSSION}

\subsection{Superconducting State}

\begin{figure}
\includegraphics[scale=0.3]{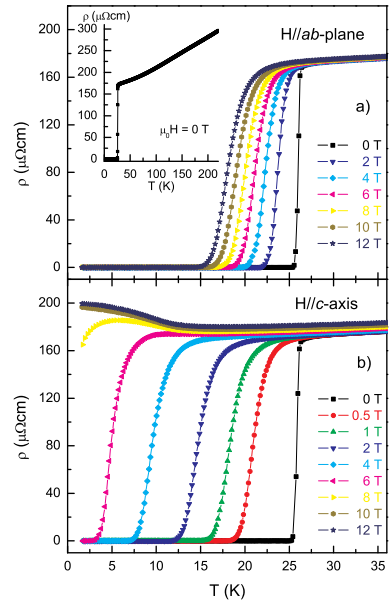}
\caption{\label{figure1} (a) Resistivity versus temperature for $H\parallel ab$. The inset shows the resistivity curve at $\mu_0H=0$ T in a wide temperature region. (b) Resistivity versus temperature for $H\parallel c$.}
\end{figure}

Figure \ref{figure1} shows the superconducting transition curves in different $H$ up to 12~Tesla for $H\parallel ab$ (Fig. \ref{figure1}a) and $H\parallel c$ (Fig. \ref{figure1}b). Upon applying the field, it is seen that the resistive transition becomes broadening for $H\parallel ab$, while for $H\parallel c$ a parallel shift of the transition is more evident. This is not unexpected by assuming a much larger $H_{c2}$ for $H\parallel ab$. The rounding of the superconducting transition in magnetic fields, however, implies that it would be difficult to accurately define the mean-field transition point $T_c(H)$ and thus to extract $H_{c2}(T)$ directly from the experimental curves, as shown in studies on hole-doped high-$T_c$ cuprates.~\cite{Welp89} In order to reliably determine $H_{c2}(T)$ of the sample, in the following we performed scaling analysis to the experimental data based on the Ginzburg$-$Landau (GL) fluctuation theory. For $H\parallel c$, it is seen that the resistivity shows an upturn at low $T$ in high fields and the $H_{c2}(0)$ could be estimated below 12~T. The low-$T$ upturn in $\rho(T)$ has been widely observed in both hole- and electron-doped cuprates near optimal-doping,~\cite{Boebinger96,Sekitani01} whose origin however has remained unclear, with localization or spin effect having been proposed.

Scaling analysis of thermodynamic and transport properties around $T_c$ has proved to be an effective way to evaluate $H_{c2}(T)$.~\cite{Welp91,Han92,Wen00,Kacmarcik04,Gao06}  For superconductors with layered-structure, Ullah and Dorsey showed that the fluctuation conductivity $\sigma_{fl}$ has a scaling form
\begin{equation}
\label{equation1}
\sigma_{fl}[\frac{H}{T}]^{1/2}=F_{2D}[\frac{T-T_c(H)}{(TH)^{1/2}}]
\end{equation}
or
\begin{equation}
\label{equation2}
\sigma_{fl}[\frac{H}{T^2}]^{1/3}=F_{3D}[\frac{T-T_c(H)}{(TH)^{2/3}}]
\end{equation}
for two-dimensional (2D) and 3D systems, respectively, with $F_{2D}$ and $F_{3D}$ the unknown scaling functions.~\cite{Ullah90,Ullah91} By using the appropriate expression to scale the experimental data, we can readily determine the parameter $T_c(H)$ with $T_c(0)$ as an additional constraint and therefore obtain the equivalent $H_{c2}(T)$.

\begin{figure}
\includegraphics[scale=0.3]{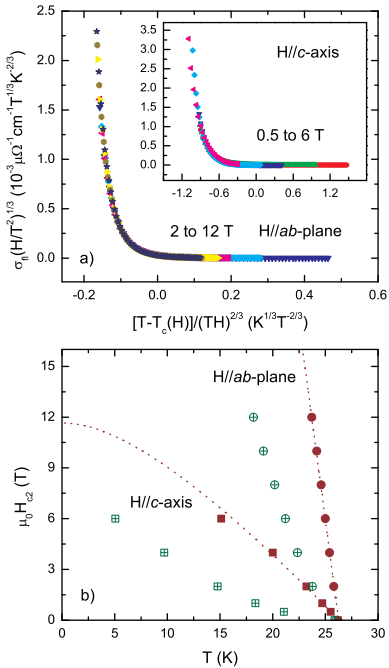}
\caption{\label{figure2} (a) 3D scaling of the fluctuation conductivity, i.e., $\sigma_{fl}(H/T^2)^{1/3}$ versus $[T-T_{c}(H)]/(TH)^{2/3}$ for $H\parallel ab$ and $2\leq\mu_0H\leq12$ T. The inset shows the same scaling analysis for $H\parallel c$ and $0.5\leq\mu_0H\leq6$ T. (b) $H_{c2}(T)$ determined from the fluctuation scaling (solid symbols) for both $H\parallel c$ and $H\parallel ab$. The dotted lines show the WHH theoretical fitting. The crossed symbols represent the points in $\rho(T)$ curves at which the resistivity becomes half the normal state value.}
\end{figure}

Figure \ref{figure2}(a) shows the scaled curves according to Eq. \ref{equation2} for both field directions. The fluctuation conductivity was obtained by subtracting the normal state conductivity ($\rho_{n}^{-1}$) from the measured conductivity, $\sigma_{fl}=\rho^{-1}-\rho_{n}^{-1}$, where $\rho_{n}$ at low $T$ was determined through an extrapolation of the normal state resistivity data between 40 and 100~K with two-order polynomial fit. For each field, the scaled data cover the resistive transition region down to temperature at which $\rho(T)$ becomes half the normal state value. As seen in Fig. \ref{figure2}(a), by adjusting the $T_c(H)$ parameter with the restriction of $T_c(0)=26.2$~K, we have obtained nice scalings of the experimental data for both $H\parallel~ab$ and $H\parallel~c$. The resultant $T_c(H)$, or equivalently $H_{c2}(T)$, were plotted as solid symbols in Fig. \ref{figure2}(b). We found that we could also obtain a 2D scaling of the data with reasonable quality according to Eq. \ref{equation1} and roughly the same $T_c(H)$, as demonstrated in a previous study in Sm$_{1.85}$Ce$_{0.15}$CuO$_{4-\delta}$ for $H\parallel c$.~\cite{Han92}

Figure \ref{figure2}(b) shows that $H_{c2}(T)$ determined from the scaling analysis exhibits linear temperature dependence in vicinity of $T_c$, as in conventional type-II superconductors. In comparison, we have also determined the points at which the resistivity becomes 50\% of the normal state value and plotted them as crossed symbols for both field directions in Fig. \ref{figure2}(b). It is seen that they exhibit positive curvature and are considerably lower than the determined $H_{c2}(T)$, similar to the observation in hole-doped cuprates.~\cite{Welp89} From fitting $H_{c2}(T)$ to the Werthamer$-$Helfand$-$Hohenburg (WHH) theory~\cite{Werthamer66} (dotted lines in Fig. \ref{figure2}(b)) we obtain $H_{c2}(0)\simeq87$~T and 11.6~T, with slope of $H_{c2}(T)$ near $T_c$ being $-$4.8~T/K and $-$0.64~T/K, for $H\parallel~ab$ and $H\parallel~c$, respectively. This indicates the in-plane coherence length $\xi_{ab}(0)\simeq53.3~{\AA}$, the $c$-axis coherence length $\xi_{c}(0)\simeq7.1~{\AA}$ and the anisotropy factor $\gamma=H_{c2}^{\parallel ab}/H_{c2}^{\parallel c}\simeq7.5$. We note that this anisotropy factor is comparable to that determined for hole-doped YBa$_2$Cu$_3$O$_{7-\delta}$ (YBCO) ($\sim5-8$ in Refs.~\onlinecite{Welp89,Nagasao08,Babic99}) but smaller than that for La$_{2-x}$Sr$_x$CuO$_4$ (LSCO) ($\sim20$ in Ref.~\onlinecite{Panagopoulos03}) and  Bi$_2$Sr$_2$CaCu$_2$O$_{8+\delta}$ (Bi2212) ($\sim60$ in Ref.~\onlinecite{Farrell89}) at optimal doping.

\subsection{Normal State}

\begin{figure}
\includegraphics[scale=0.3]{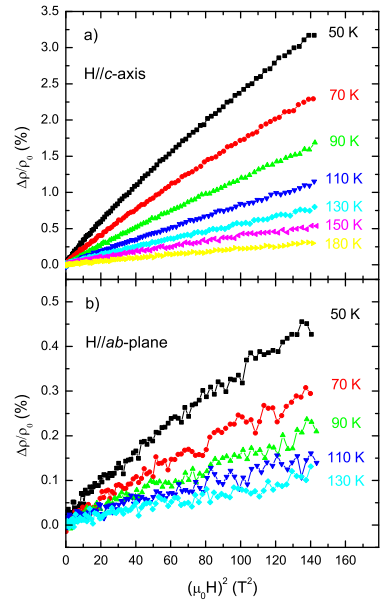}
\caption{\label{figure3} Normal state MR versus $H^2$ for $H\parallel c$ (a) and $H\parallel ab$ (b).}
\end{figure}

Now we turn to the normal state above $T_c$. Figure \ref{figure3} shows the MR $\Delta\rho/\rho_{0}$~($\Delta\rho=\rho_{H}-\rho_{0}$ with $\rho_{H}$ and $\rho_{0}$ the resistivity in field $H$ and in zero field respectively) of the crystal at different temperatures ($T\geq50$~K), in the plot of $\Delta\rho/\rho_{0}$ vs. $H^2$. For both field directions the positive MR shows conventional orbital MR behavior in the weak-field regime,~\cite{Ziman72} that is, it basically follows an $H^2$ dependence and its strength decreases with increasing temperature. At lower $T$ (50 and 70~K), small deviation from the $H^2$ behavior may come from magnetic-field suppression of superconducting fluctuations.~\cite{Ando02} For $H\parallel~c$, the MR is rather large with the order of one percent, similar to previous reports.~\cite{Gollnik98} This is contrasted with what we observed in hole-doped YBCO or LSCO single crystals, where the normal state MR for $H\parallel~c$ was about one order of magnitude smaller at similar temperature and field range.~\cite{Kimura96,Harris94} In hole-doped cuprates the normal state MR in transverse configuration is usually ascribed to the orbital scattering within a single band picture. Whereas, in NCCO and other electron-doped cuprates, the large MR, together with other physical properties such as Hall effect, has been widely interpreted as an indiction of two-band transport.~\cite{Seng95,Gollnik98} According to classical transport theory, the orbital MR could be enhanced when different types of charge carriers participating in the electrical conduction.~\cite{Ziman72}

\begin{figure}
\includegraphics[scale=0.3]{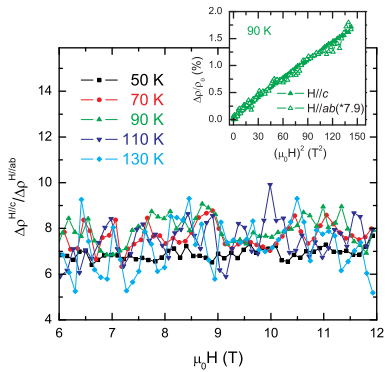}
\caption{\label{figure4} The ratio of the MR between $H\parallel c$ and $H\parallel ab$ as a function of $H$. The inset shows the comparison of the MR for both field directions at 90 K with data for $H\parallel ab$ multiplied by 7.9.}
\end{figure}

It is seen in Fig. \ref{figure3} that the MR for $H\parallel~ab$ is considerably smaller than that for  $H\parallel~c$ at the same temperature and field scale, which is similar to the observation in hole-doped LSCO single crystals.~\cite{Kimura96} This anisotropy provides an additional evidence that the measured MR with $H\parallel c$ is dominated by the orbital contribution. As we know, if the MR mainly originated from the field coupling to spin degree of charge carriers, it would be almost isotropic. It is worth mentioning that here we have not considered the possible effect of field-induced changes in sample's spin structure on the MR. In lightly doped Pr$_{1.3-x}$La$_{0.7}$Ce$_x$CuO$_4$ (PLCCO, $x=0.01$) and NCCO ($x=0.025$), it was reported that field-induced spin-flop transitions in the spin structure resulted a much larger, distinct field-dependent MR with $H\parallel~ab$ at low $T$.~\cite{Lavrov04,Li05} Figure \ref{figure4} shows the anisotropy of the MR more explicitly, by plotting the ratio $\zeta=\Delta\rho^{H\parallel~c}/\Delta\rho^{H\parallel~ab}$ as a function of $H$. $\zeta$ is around 7 and nearly independent on $H$ and $T$ in present experiment. Inset of Fig. \ref{figure4} shows the MR at 90~K as an example, which was plotted as a function of $H^2$. When the MR for $H\parallel~ab$ is multiplied by 7.9, we can see it follows nearly the same line as the MR for $H\parallel~c$.

It may be noted that the normal state MR ratio $\zeta$ is close to the anisotropy ratio $\gamma$ determined above for the superconducting state. On the one hand, in our view the closeness of both parameters may be merely coincidental and have no obvious physical importance. On the other hand, however, we point out that there should be an internal connection between $\zeta$ and $\gamma$, since both parameters would relate to the anisotropy of the effective mass $m^\ast$ of charge carriers. In anisotropic GL theory, $\gamma=H_{c2}^{\parallel ab}/H_{c2}^{\parallel c}=\sqrt{m_c^\ast/m_{ab}^\ast}$ with $m_c^\ast$ and $m_{ab}^{\ast}$ the effective mass along $c$-axis and within $ab$-plane respectively. This implies that the axial effective mass $m_c^\ast$ is about 50 times heavier than the in-plane $m_{ab}^\ast$ for our NCCO crystal. For the normal state MR, as indicated in the two-band model, it is governed by the $m^\ast$, the scattering rate $\tau$ and other properties of the carriers.~\cite{Ziman72} The anisotropy of the $m^\ast$ thus would certainly contribute to the anisotropy of the MR, namely, the ratio $\zeta$, though a simple relation between them is difficult to obtain since the MR is determined by aforementioned parameters in a complicated way, especially with the presence of different types of carriers. Nevertheless, the present study shows that the optimally doped NCCO single crystal has a moderate anisotropy in both superconducting and normal state.

\section{CONCLUSION}

In summary, by investigating the upper critical field $H_{c2}$ and the normal state MR with field $H$ either parallel or perpendicular to the crystallographic $c$-axis, we have determined the anisotropy properties of an optimally doped NCCO single crystal. $H_{c2}$ estimated from scaling of the fluctuation conductivity is about 87~T and 11.6~T for $H\parallel~ab$ and $H\parallel~c$ respectively, which yields $\xi_{ab}(0)\simeq53.3~{\AA}$, $\xi_{c}(0)\simeq7.1~{\AA}$ and the anisotropy factor $\gamma=H_{c2}^{\parallel ab}/H_{c2}^{\parallel c}\simeq7.5$. The normal state MR for $H\parallel~ab$ is found to be almost a magnitude smaller than that for $H\parallel~c$. This anisotropy, together with the $H^2$ dependence, confirms that the MR with $H\parallel~c$ is mainly due to the orbital scattering. Present findings place optimally doped NCCO cuprate as an anisotropic 3D superconductor with a moderate anisotropy.

\begin{acknowledgments}

We are grateful to Profs. S. L. Li and P. Dai for providing the NCCO single crystal and for helpful comments. We are also indebted to Prof. H. H. Wen for experimental support and helpful discussions.

\end{acknowledgments}

\bibliography{NCCOAnisotropy}

\end{document}